\documentclass[10pt,prx,twocolumn,superscriptaddress,aps]{revtex4-2}

\usepackage{graphicx}
\usepackage{dcolumn}
\usepackage{bm}
\usepackage{amsmath}
\usepackage{amsfonts}
\usepackage{amssymb}
\usepackage{dcolumn}
\usepackage{epsfig}
\usepackage{graphicx}
\usepackage{latexsym}
\usepackage{setspace}
\usepackage{fancyhdr}
\usepackage{wrapfig}
\usepackage{textgreek}
\usepackage{ulem}
\usepackage{titlesec}
\usepackage{outlines}
\usepackage{multirow}
\usepackage{adjustbox}
\usepackage{enumitem}
\usepackage[official]{eurosym}
\usepackage{mdframed}
\usepackage{enumitem}
\usepackage{braket}
\usepackage{hhline}
\usepackage[title]{appendix}
\usepackage[dvipsnames]{xcolor}
\usepackage{comment}
\usepackage{titlesec}
\usepackage{lipsum}
\usepackage{upgreek}
\titlespacing{\section}{0pt}{1.5em}{1.5em}
\titlespacing{\subsection}{0pt}{1.5em}{1.5em}

\begin{document}



\title{A Wafer-Scale Heterogeneous III-V-on-Silicon Nitride Quantum Photonic Platform}
\author{Lillian Thiel}
\affiliation{\protect\hbox{Electrical and Computer Engineering Department, University of California, Santa Barbara, CA, USA}}
\author{Boqiang Shen}
\affiliation{\protect\hbox{Nexus Photonics Inc., Goleta, CA 93117, USA}}
\author{Jasper R. Venneberg}
\affiliation{\protect\hbox{LIGO, Massachusetts Institute of Technology, Cambridge, Massachusetts, USA}}
\author{Melissa A. Guidry}
\affiliation{\protect\hbox{LIGO, Massachusetts Institute of Technology, Cambridge, Massachusetts, USA}}
\author{Nic Arnaud}
\affiliation{\protect\hbox{Department of Electrical and Computer Engineering, University of Virginia, Charlottesville, VA, USA}}
\author{Adam Slater}
\affiliation{\protect\hbox{Department of Electrical and Computer Engineering, University of Virginia, Charlottesville, VA, USA}}
\author{Lucas Wang}
\affiliation{\protect\hbox{Physics Department, University of California, Santa Barbara, CA, USA}}
\author{Xuefeng Li}
\affiliation{\protect\hbox{Electrical and Computer Engineering Department, University of California, Santa Barbara, CA, USA}}
\author{Josh Castro}
\affiliation{\protect\hbox{Electrical and Computer Engineering Department, University of California, Santa Barbara, CA, USA}}
\author{Yiming Pang}
\affiliation{\protect\hbox{Electrical and Computer Engineering Department, University of California, Santa Barbara, CA, USA}}
\author{Max Meunier}
\affiliation{\protect\hbox{Electrical and Computer Engineering Department, University of California, Santa Barbara, CA, USA}}
\author{Sahil D. Patel}
\affiliation{\protect\hbox{Electrical and Computer Engineering Department, University of California, Santa Barbara, CA, USA}}
\author{Yang Shen}
\affiliation{\protect\hbox{Nexus Photonics Inc., Goleta, CA 93117, USA}}
\author{Theodore Morin}
\affiliation{\protect\hbox{Nexus Photonics Inc., Goleta, CA 93117, USA}}
\author{Igor Kudelin}
\affiliation{\protect\hbox{Nexus Photonics Inc., Goleta, CA 93117, USA}}
\author{Bowen Song}
\affiliation{\protect\hbox{Nexus Photonics Inc., Goleta, CA 93117, USA}}
\author{Kaustubh Asawa}
\affiliation{\protect\hbox{Nexus Photonics Inc., Goleta, CA 93117, USA}}
\author{John E. Bowers}
\affiliation{\protect\hbox{Electrical and Computer Engineering Department, University of California, Santa Barbara, CA, USA}}
\affiliation{Materials Department, University of California, Santa Barbara, CA, USA}
\author{Kerry Vahala}
\affiliation{\protect\hbox{T.J. Watson Laboratory of Applied Physics, California Institute of Technology, Pasadena, CA, USA}}
\author{Nergis Mavalvala}
\affiliation{\protect\hbox{LIGO, Massachusetts Institute of Technology, Cambridge, Massachusetts, USA}}
\author{Xinghui Yin}
\affiliation{\protect\hbox{LIGO, Massachusetts Institute of Technology, Cambridge, Massachusetts, USA}}
\author{Steven Bowers}
\affiliation{\protect\hbox{Department of Electrical and Computer Engineering, University of Virginia, Charlottesville, VA, USA}}
\author{Minh A. Tran}
\affiliation{\protect\hbox{Nexus Photonics Inc., Goleta, CA 93117, USA}}
\author{Tin Komljenovic}
\affiliation{\protect\hbox{Nexus Photonics Inc., Goleta, CA 93117, USA}}
\author{Galan Moody}
\email{moody@ucsb.edu}
\affiliation{\protect\hbox{Electrical and Computer Engineering Department, University of California, Santa Barbara, CA, USA}}


\begin{abstract}
\noindent Heterogeneous integration of gain and strongly nonlinear materials with ultra-low-loss silicon nitride (SiN) photonics offers a route to scalable quantum circuits, but concurrent wafer-scale manufacturability, low interlayer loss, and high performance have been challenging to realize. Here we demonstrate a wafer-scale III-V-on-SiN quantum photonic platform that directly integrates III-V layers to foundry-fabricated SiN circuits. The SiN layer provides 200-300 nm thick waveguides with $<1$~dB/m loss and a mature passive photonics ecosystem, while III-V materials provide large $\chi^{\left(2\right)}$ and $\chi^{\left(3\right)}$ nonlinearities for parametric gain, frequency conversion and quantum light generation. Adiabatic interlayer couplers yield $<25$~mdB loss to InGaP waveguides and resonators with intrinsic quality factors exceeding $10^6$, enabling $15\times$ brighter entanglement sources and efficient nonlinear conversion on SiN. Integrated components--including low-loss beam splitters, waveguide crossers, and tunable interferometers--are complemented by III–V lasers and InP photodetectors with amplifiers achieving up to $99^{+1}_{-12}$~\% quantum efficiency and $3$~GHz bandwidth. This architecture unites ultra-efficient sources, nonlinear elements and detectors on a wafer-scale, low-loss platform, establishing a path toward large-scale, low-noise quantum photonic systems.

\end{abstract}
\maketitle

\thispagestyle{plain}


\section*{Introduction}

\noindent Silicon nitride (SiN) integrated photonics has emerged as a leading platform for quantum technologies due to its exceptional linear optical properties, offering sub-dB/m transmission, broad transparency, high phase stability, and compatibility with high-volume semiconductor manufacturing \cite{bauters2011ultra, roeloffzen2018low,rahim2017expanding,blumenthal2018silicon,ji2023ultra}. These attributes have enabled the development of increasingly complex passive circuits for quantum state manipulation and routing \cite{taballione20198,li2025down,chen2024ultralow,aghaee2025scaling}. While SiN excels as a low-loss passive platform, its lack of intrinsic optical gain and modest second- and third-order nonlinearities \cite{boyd2008nonlinear} motivate hybrid or heterogeneous integration with complementary materials offering strong nonlinearities, optical gain, and tight confinement, while preserving wafer-scale manufacturability for quantum photonics.

III-V semiconductors provide a natural complement to SiN, combining direct bandgap gain media with some of the largest $\chi^{\left(2\right)}$ and $\chi^{\left(3\right)}$ nonlinearities available in any integrated photonic material \cite{baboux2023nonlinear,chang2022csoi,mobini2022algaas}. Among them, InGaP is particularly attractive due to its wide bandgap, absence of two-photon absorption at telecommunications wavelengths, and ability to support sub dB-cm waveguides \cite{ahler2026low}, efficient frequency converters \cite{hu2026efficient}, and entanglement sources \cite{thiel2024wafer,akin2024ingap}. Yet integrating III-V materials with SiN at wafer scale has proven challenging. Prior demonstrations have typically relied on adhesive layers or complex pattern-transfer processes that limit scalability, introduce optical loss, or restrict device layout flexibility\cite{OpdeBeeck:20, Xiang:22}. Furthermore, achieving low interlayer coupling loss, which is critical for routing light efficiently between nonlinear, gain, and passive components, has remained a central obstacle.

\begin{figure*}[t!]
    \centering
    \includegraphics[width=\textwidth]{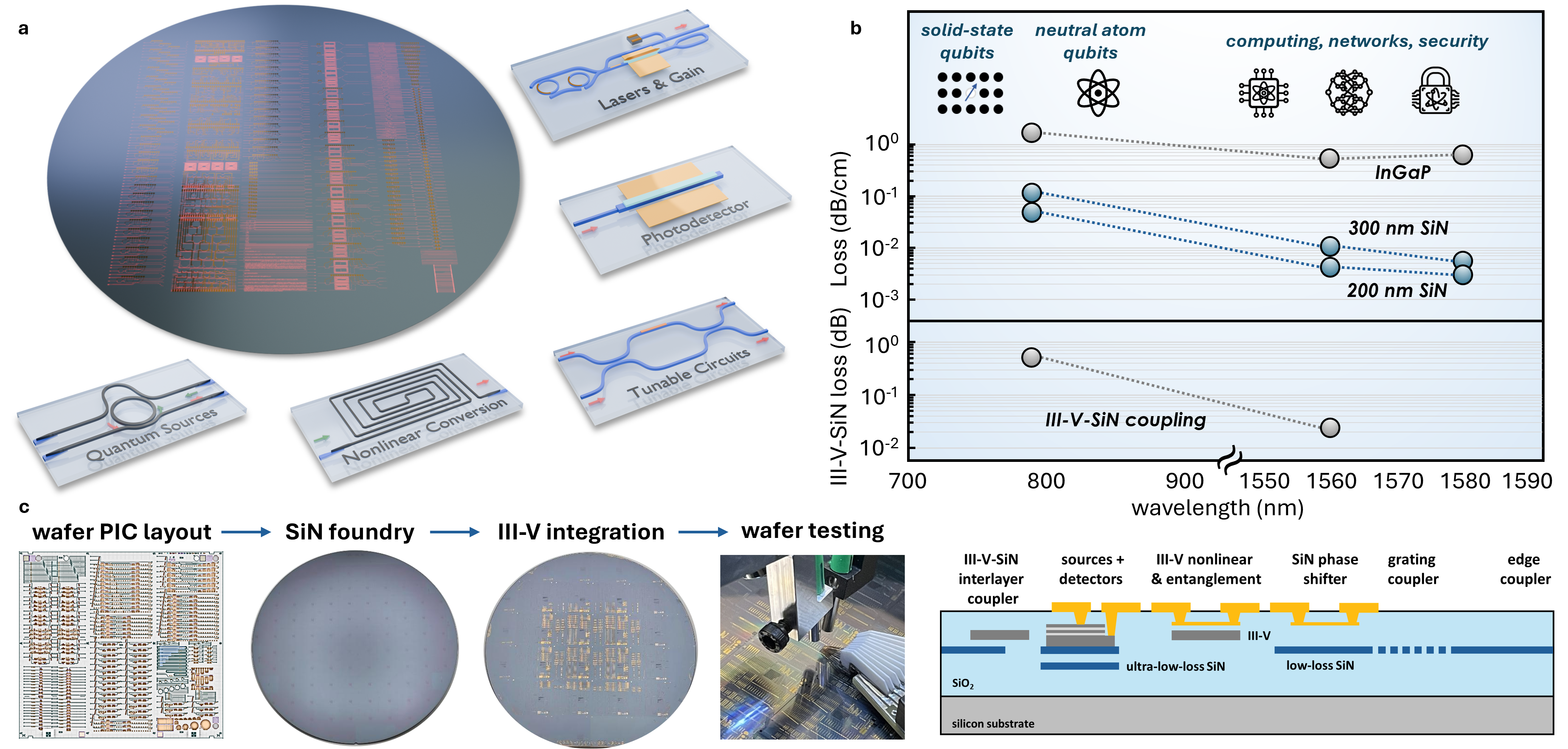}
    \vspace{-15pt}
    \caption{\small \label{Fig1} \textbf{Heterogeneous III-V-on-SiN platform.} \textbf{(a) }Illustration of our scalable nonlinear quantum photonics platform, which includes quantum entanglement sources, nonlinear frequency conversion and generation, tunable interferometric circuits, high-efficiency integrated photodetectors, and tunable low-noise semiconductor lasers. \textbf{(b)} Measured waveguide propagation losses for the III-V InGaP layer and the two SiN layers, and the measured III-V-to-SiN adiabatic layer transition coupling losses near 1560 nm and 780 nm \textbf{(c)} Wafer fabrication flow from internal modeling, design, and layout to SiN foundry fabrication, back-end-of-line III-V integration, and wafer-scale testing. The right panel illustrates the cross sections of different device components on the platform.}
    \vspace{-10pt}
\end{figure*}

Here we demonstrate a wafer-scale III-V-on-SiN quantum photonic platform that overcomes these longstanding challenges by combining direct bonding of III-V materials with foundry-fabricated ultra-low-loss SiN photonic integrated circuits (PICs). Figure \ref{Fig1} illustrates our wafer-scale manufacturing and testing process. Our platform enables seamless integration of high-nonlinearity, gain elements, and photodetection components with ultra-low-loss dielectric waveguides. We have developed adiabatic interlayer couplers that achieve coupling losses below 25~mdB at 1560~nm and InGaP waveguide losses below 1~dB/cm (8~dB/cm) in the telecom C-band (NIR). High-Q InGaP microresonators, low-loss tunable SiN components, and integrated III-V lasers and photodetectors onto the SiN platform collectively enable a level of performance and functionality not previously achieved in quantum photonics. This architecture provides an extensible foundation for large-scale, low-noise quantum photonic systems incorporating ultra-efficient sources, nonlinear devices, and quantum frequency converters on a common wafer-scale platform. 

\section*{III-V Nonlinear Frequency Generation and Quantum Sources on SiN}

\begin{figure*}[t!]
    \centering
    \includegraphics[width=\textwidth]{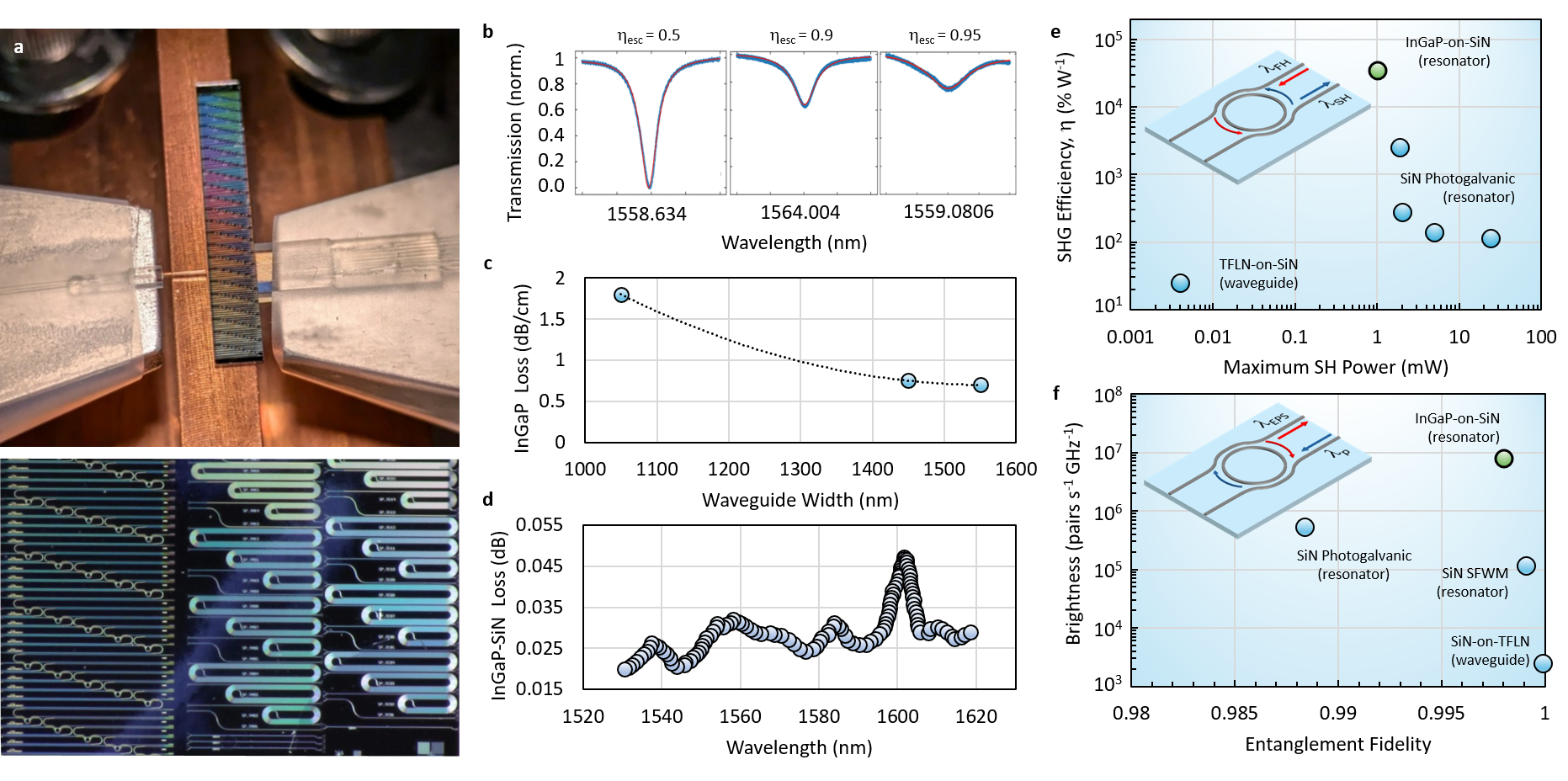}
    \vspace{-15pt}
    \caption{\small \label{Fig2} \textbf{InGaP nonlinear frequency generation and entanglement sources on SiN}. \textbf{(a)} Images of an InGaP-on-SiN singulated chip under test, including (bottom) add-drop microresonators and up to 10-cm-long spiral waveguides. \textbf{(b)} Representative microresonator transmission spectra near 1560 nm for resonators designed for escape efficiencies of $\eta_{esc} =$ 0.5, 0.9, and 0.95 (left-to-right). \textbf{(c)} InGaP propagation loss extracted from microresonator transmission data near 1560 nm versus ring width, demonstrating loss as low as 0.7 dB/cm. \textbf{(d)} Measured III-V-to-SiN adiabatic transition losses for 1550 nm TE modes, demonstrating $<25$ mdB coupling loss at 1560 nm. \textbf{(e)} Second harmonic generation (SHG) efficiency, $\eta$, versus maximum on-chip SHG pump power on SiN for various configurations, including thin film lithium niobate (TFLN)-on-SiN~\cite{Vand2023}, photogalvanic-poled SiN~\cite{Lu2021, Li2023, Clementi2023}, and our results with InGaP-on-SiN (green circles). \textbf{(f)} Measured entangled-photon pair brightness versus entanglement fidelity demonstrated on SiN. Data are normalized to 10~$\upmu$W on-chip pump power for comparison.}
    \vspace{-10pt}
\end{figure*}

Figure \ref{Fig2} illustrates the InGaP-on-SiN chiplets and their performance for nonlinear frequency generation and entangled-photon pair sources. The fabrication process is detailed in the supplementary information. After SiO$_2$-clad 200-mm SiN wafers are received from the foundry, they are cored and planarized, leaving a $\sim50$~nm SiO$_2$ spacer between the 200-nm thick SiN and 105-nm thick InGaP layers. InGaP is direct bonded through an established process \cite{thiel2024wafer, li2026highefficiencyingap}, followed by PECVD cladding and metallization for thermal tuning. The top panel of Fig. \ref{Fig2}a shows a chiplet with hundreds of add-drop resonator entangled-pair sources, and the bottom panel is a zoom-in of spiral waveguides for loss cut-back measurements and resonator arrays for spontaneous parametric down conversion (SPDC) and second harmonic generation (SHG). 

Linear transmission spectra are shown in Fig. \ref{Fig2}b for three representative InGaP resonators with different waveguide bus-resonator pulley coupler designs corresponding to critically coupled (escape efficiency $\eta_{esc} = 0.5$), overcoupled ($\eta_{esc} = 0.9$), and strongly overcoupled ($\eta_{esc} = 0.95$) near 1560~nm resonances. The average propagation loss extracted from fitting resonator lineshapes varies from $<2$~dB/cm for resonator widths of 1.05~$\upmu$m to $\sim0.65$~dB/cm for widths of 1.50~$\upmu$m, as shown in Fig. \ref{Fig2}c. Likewise, at 780~nm loss cut-back measurements demonstrate waveguide propagation losses of 8~dB/cm. Additionally, the InGaP fabrication process shows no significant impact on propagation losses in the SiN waveguide layer. Measurements and analysis of the propagation losses and phase-matching conditions are presented in the supplementary information.

These InGaP-on-SiN wafers are designed for efficient InGaP-SiN interlayer coupling near 1560~nm. Adiabatic inverse tapers are designed for efficiently transferring C-band fundamental transverse electric (TE) modes between the two layers. Cut-back measurements across the wafer indicate interlayer coupling losses below 0.05~dB (98.9~$\%$ transmission) across the entire C-band and as low as 0.02~dB (99.5~$\%$ transmission), shown in Fig. \ref{Fig2}d. Adiabatic inverse tapers are also designed in the same InGaP-SiN layer for coupling 780~nm fundamental transverse magnetic (TM) modes, which is required to achieve modal phase matching for the SPDC and SHG processes in the waveguides and resonators \cite{Fontaine2025, thiel2024wafer}. In the 780~nm wavelength band, $\sim0.5$~dB (90~$\%$ transmission) couplers are demonstrated (supplementary information), which can be improved using thicker SiN or smaller inverse taper tips.

Figures \ref{Fig2}e,f demonstrate the key capabilities enabled by integrating InGaP onto SiN. In Fig. \ref{Fig2}e, the normalized SHG efficiency versus maximum measured on-chip SHG power is shown for different approaches (blue symbols) and our InGaP-on-SiN platform (green symbol). For waveguides, we have previously demonstrated efficiencies up to $5000~\%$ W$^{-1}$, absolute conversion efficiencies above $61~\%$, and maximum on-chip powers exceeding $50$~mW \cite{ahler2026low}. Here, we demonstrate SHG in high-Q InGaP-on-SiN resonators, where saturation occurs at lower fundamental pump powers, but the doubly resonant structures enable SHG efficiencies exceeding 35,000~$\%$ W$^{-1}$. These are the highest reported SHG efficiencies in SiN photonics to the best of our knowledge. 

Add-drop InGaP microresonators are also designed for entangled-photon pair generation via SPDC. The 780 nm (1560 nm) pulley couplers are designed for critical (over) coupling. Devices with various resonator and waveguide widths and resonator radii have been studied and presented in the supplementary information. The resonators are pumped on resonance near 780 nm, and photon pairs are coupled off chip and measured with superconducting nanowire single-photon detectors near 1560 nm to characterize the entangled-photon pair brightness (on-chip generated pairs s$^{-1}$ GHz$^{-1}$ bandwidth) and the coincidence-to-accidental ratio (CAR), which is used to estimate the entanglement fidelity. Figure \ref{Fig2}f shows a summary of these measurements, where the blue symbols indicate prior studies of pair generation in SiN or SiN-on-thin film lithium niobate\cite{Vand2023, Lu2021, Li2023, Clementi2023}. The green symbol highlights pair generation from our platform, whereby InGaP integration enables brightness of nearly $10^{8}$ pairs s$^{-1}$ GHz$^{-1}$ (normalized to 10 $\upmu$W pump power), CAR higher than $10^4$, and two-photon entanglement fidelities of $99.9\%$ at $10^6$ on-chip pairs s$^{-1}$. This brightness is $>15\times$ higher than any native SiN sources. Details of the SHG and SPDC devices, setups, and characterization are included in the supplementary information.

\begin{figure*}[t!]
    \centering
    \includegraphics[width=\textwidth]{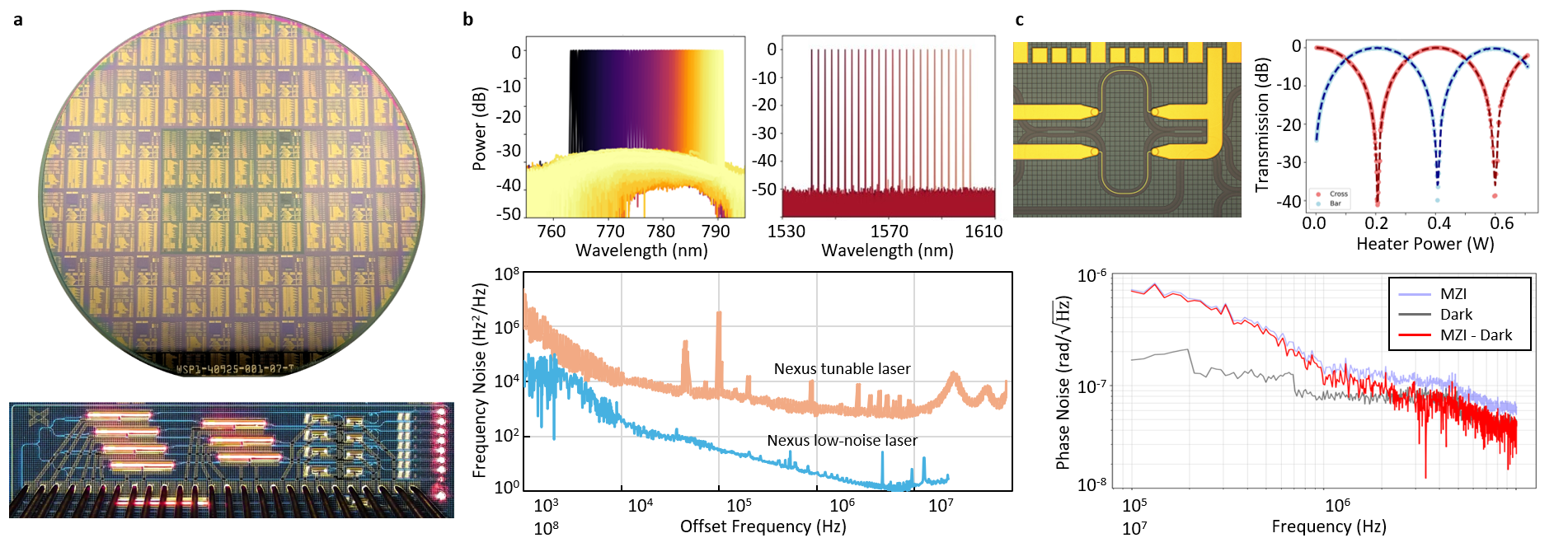}
    \vspace{-15pt}
    \caption{\small \label{Fig3} \textbf{Low-Noise and tunable integrated lasers and interferometers.} \textbf{(a)} Fabricated SiN wafer with integrated III-V semiconductor lasers and amplifiers, with a multi-channel tunable semiconductor amplifier chip shown in the bottom panel. \textbf{(b)} Both 1560 nm and 780 nm tunable lasers are demonstrated with kilohertz (hertz) level linewidths without (with) injection locking, side-mode suppression ratios as high as 50 dB, and off-chip output power exceeding 10 mW are demonstrated. \textbf{(c)} Integrated Mach Zehnder interferometer with $<6$~mdB insertion loss, $>40$~dB extinction, and $<0.4$~mrad interferometric phase noise (without locking) integrated across 100~kHz to 10~MHz.}
    \vspace{-10pt}
\end{figure*}

\section*{Active III-Vs on Low-Loss SiN}

The absence of on-chip optical gain and coherent light sources remains a fundamental limitation of purely passive SiN quantum photonic circuits. Reliance on external lasers introduces excess coupling loss, increased phase noise, and limits scalability as circuit complexity grows. To overcome these constraints, we have heterogeneously integrated active III-V devices directly onto SiN waveguides, enabling wafer-scale realization of tunable semiconductor lasers, semiconductor optical amplifiers (SOAs), and monitor photodiodes co-integrated with passive circuitry \cite{tran2022extending, zhang2023photonic}. As shown in Figure \ref{Fig3}a, the platform is fabricated by bonding III-V epitaxial stacks onto patterned SiN wafers originally fabricated at the 200-mm CMOS foundry scale and subsequently cored to 100 mm for back-end heterogeneous processing. All active components are defined in the bonded III-V layers and optically coupled to the underlying SiN waveguides, forming a unified active-passive photonic platform suitable for large-scale quantum and classical photonic circuits. As an illustrative example, the photonic integrated circuit shown in Figure \ref{Fig3}a incorporates a single seed laser that branches into seven co-integrated semiconductor optical amplifiers (SOAs), each equipped with monitor photodiodes. This on-chip gain integration enables scalable power distribution and dynamic signal management within complex photonic circuits, while preserving the low-loss and phase-stable characteristics of the underlying SiN platform.

Tunable semiconductor lasers operating at 1560 nm and 780 nm are implemented based on wavelength-selective SiN Vernier ring resonator filters embedded within extended laser cavities. Two microring resonators with slightly detuned free spectral ranges form a compact, thermally tunable filter mirror that enforces single-mode operation when combined with III-V gain sections, following architectures previously demonstrated on this platform \cite{tran2022extending}. Using this approach, broadly tunable lasers with tuning ranges exceeding 60 nm at 1560 nm and 25 nm at 780 nm are demonstrated, while maintaining side-mode suppression ratios greater than 40 dB across the full tuning ranges (Figure \ref{Fig3}b). Direct III-V emission is used at both wavelength bands, enabled by InP-based epitaxy for 1560 nm and GaAs-based epitaxy for 780 nm, providing flexible access to both telecommunications and visible/near-visible spectral regions within the same heterogeneous photonic framework.

The low-loss SiN cavity and filter structures enable multiple grades of laser coherence tailored to different system requirements. Free-running tunable lasers employing compact Vernier filters exhibit fundamental linewidths in the kilohertz regime, benefiting from the long effective cavity lengths and reduced optical loss provided by SiN waveguides. For applications requiring ultra-high coherence, laser frequency noise is further suppressed via self-injection locking to ultra-high-Q SiN resonators implemented here in a multilayer SiN architecture, achieving linewidths in the hertz-level regime, as evidenced by frequency noise measurements shown in Figure \ref{Fig3}b.

\begin{figure*}[t!]
    \centering
    \includegraphics[width=\textwidth]{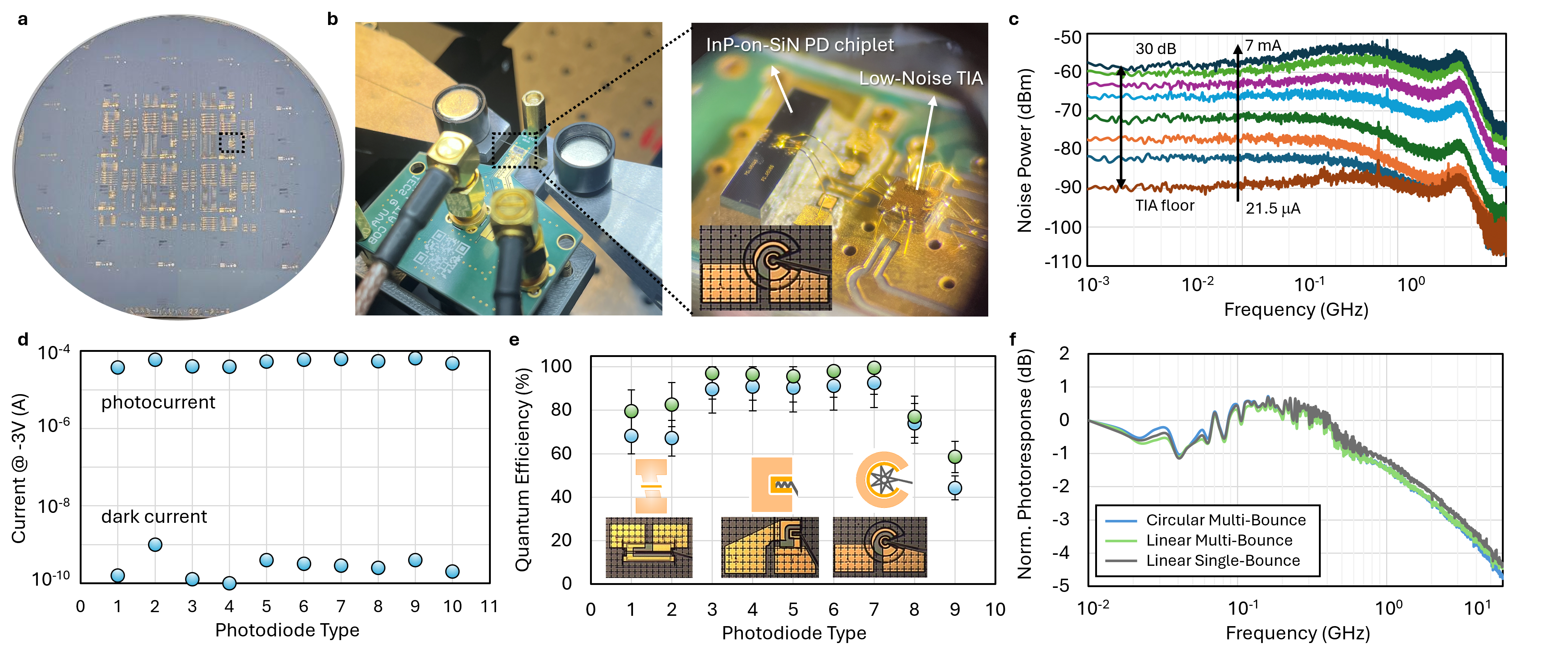}
    \vspace{-15pt}
    \caption{\small \label{Fig4} \textbf{High quantum efficiency and low-noise photodetectors.} \textbf{(a)} Image of completed InP-on-SiN photodetector wafer. \textbf{(b)} Packaged photodetector chip with low-noise transimpedance amplifier. \textbf{(c)} Electronic noise clearance at 1 mW on-chip optical pump power at 1550 nm and 3-dB bandwidth of the packaged detector and amplifier. \textbf{(d-f)} Dark current, photocurrent, and calibrated quantum efficiency of the different photodetector configurations. Quantum efficiency exceeding 99$\%$ is measured for the circular multi-bounce designs. Photodiode Types 1-2: single-bounce reference designs; 3-6: linear multi-bounce designs; 7-9: circular multi-bounce designs. The green and blue points represent nominally identical photodiodes measured from two separate chips from the same wafer.}
    \vspace{-10pt}
\end{figure*}

In addition to active light sources, we also realize SiN Mach-Zehnder interferometers (MZIs) integrated on the same platform (Figure \ref{Fig3}c). The MZIs employ balanced arm lengths and symmetric directional couplers designed for 50:50 splitting at 1560 nm, with integrated microheaters providing precise thermo-optic phase tuning. Owing to the ultra-low propagation loss of the SiN waveguides and careful coupler optimization, device-level insertion losses below 10 mdB and extinction ratios exceeding 40 dB (measurement limited) are achieved. Interferometric phase stability measurements performed without active locking show RMS phase noise $\theta_\text{RMS}<0.4$~mrad (integrated across 100~kHz to 10~MHz), which sets an upper bound on the common-mode rejection ratio (CMRR) of $-62$~dB due to fluctuations around mid-fringe during measurement. Together with the integrated low-noise lasers, these MZIs provide essential building blocks for scalable photonic systems requiring precise phase control, coherent routing, and stable interferometric operation.

\section*{High Efficiency and Low-Noise Photodetector Integration}

High-efficiency on-chip photodetection is a critical requirement for scalable quantum photonic systems \cite{tasker2024bi,lita2022development,moody20222022}, where photodetection efficiency is especially critical for quantum applications requiring homodyne detection including sensing using squeezed light \cite{lawrie2019quantum}, continuous-variable cluster states for quantum computing \cite{larsen2019deterministic,asavanant2019generation}, and continuous-variable quantum communication \cite{lodewyck2007quantum,pirandola2020advances}. To address this need, we optimized our design for InP-on-SiN photodetector for near-unity quantum efficiency at 1550~nm, fabricated using the same heterogeneous III-V/SiN integration process as the active laser platform. As shown in Figure \ref{Fig4}a, the wafer comprises a large number of photodiodes integrated on SiN waveguides, enabling systematic exploration of different detector geometries. The platform targets telecom-band detection, which is central to the quantum architectures demonstrated in this work, where entangled photons are generated and routed at 1550~nm. The same design concepts are directly extensible to shorter wavelengths, including 775~nm, using appropriate III-V absorber materials.

A key feature of this platform is the photodetector coupling and absorption strategy, which builds on the unique III-V/SiN mode-transfer approach developed for this heterogeneous system. Light propagating in the SiN waveguide is first evanescently coupled into an intermediary dielectric waveguide and then butt-coupled into the III-V absorber region. In contrast to conventional waveguide-integrated photodiodes, once light enters the absorber region it is no longer confined to a guided mode, but instead propagates in a quasi-free-space regime within the III-V material. This enables photodetector designs that suppress photon loss at the III-V interface and deliberately exploit multiple reflections within the absorber to enhance absorption probability. Figures~\ref{Fig4}d--f summarize three classes of photodetector geometries explored in this work. Photodiode Types~1--2 serve as single-bounce reference designs, in which the intermediary waveguide is angled relative to the absorber interface to suppress back-reflection into the SiN waveguide, at the expense of an intrinsic photon loss associated with return loss at the III-V interface. Photodiode Types~3--6 implement linear multi-bounce geometries, where the absorber sidewalls are shaped to confine the incoming light over several reflections, enabling near-unity absorption with only a few passes, balanced against diffraction-induced beam spreading. Photodiode Types~7--9 employ circular multi-bounce geometries that confine light within a compact, rotationally symmetric absorber region, effectively forming an integrated analog of an optical integrating-sphere detector. This geometry allows photons to undergo many reflections within the absorber, dramatically increasing absorption probability while maintaining a compact footprint. Across these designs, the circular multi-bounce photodiodes achieve the highest absorption efficiency and form the basis for the near-unity quantum efficiency reported here.

\begin{figure*}[t!]
    \centering
    \includegraphics[width=0.9\textwidth]{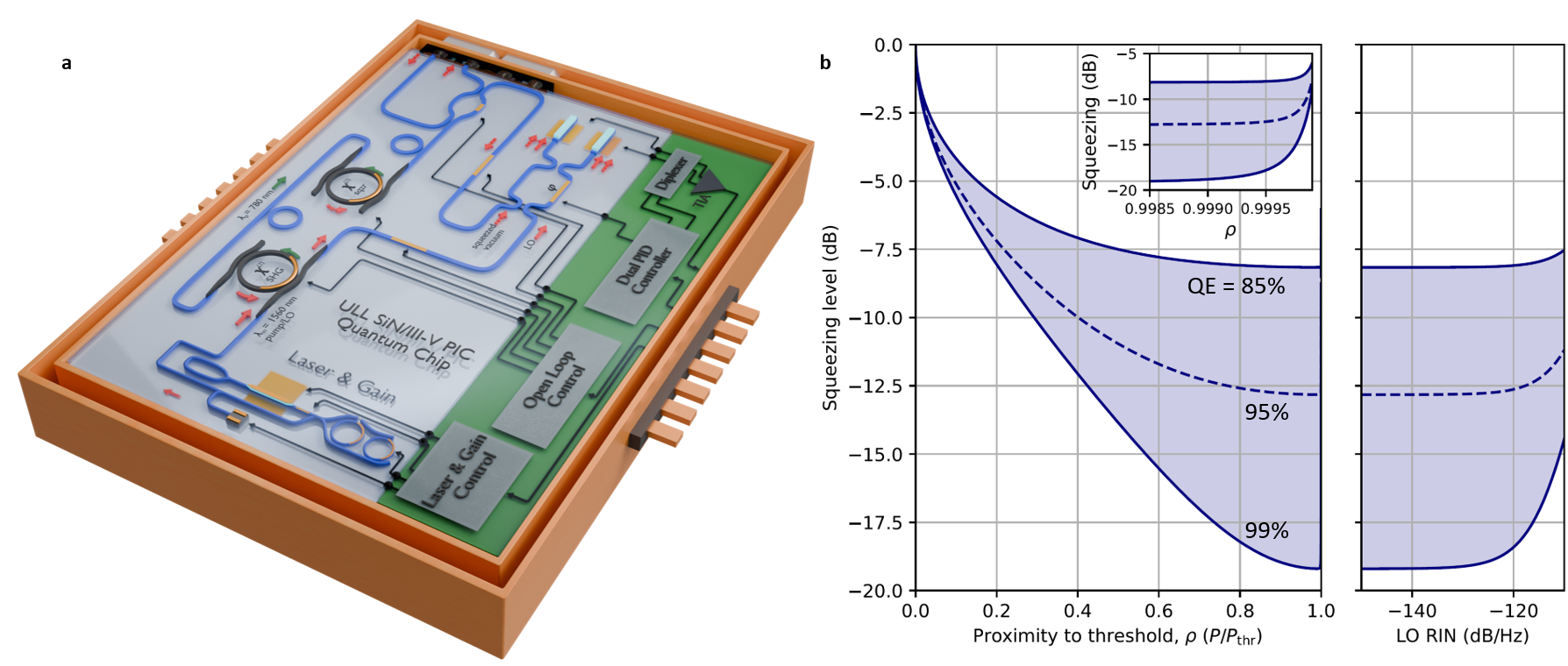}
    \vspace{-5pt}
    \caption{\small \label{Fig5} \textbf{Concept of fully integrated quantum photonic transceiver.} \textbf{(a)} Illustration of a co-packaged photonic-electronic device for squeezed light generation, efficient off-chip transmission and collection, and on-chip detection. The base platform comprises SiN waveguides, filters, and a tunable interferometer, with integrated InGaP SHG and squeezing stages and integrated InP gain and balanced photodetection. \textbf{(b)} Maximum measurable squeezing considering only laser noise, detection path noise, and losses. Left: Maximum measurable squeezing versus proximity to the optical parametric oscillation threshold, given as the ratio of the in-waveguide pump power over the threshold power. The solid line bounds correspond to the photodiode quantum efficiencies (QE) of $85$-$99$\%, where the dashed line corresponds to $95$\%. The model includes photodetector dark noise, laser frequency noise, and local oscillator (LO) relative intensity noise (RIN) with CMRR = $-62$~dB.  Right: Assuming proximity to threshold of $0.99$, the maximum measurable squeezing level versus LO RIN. }
    \vspace{-10pt}
\end{figure*}

Electrical and optical characterization of the photodetectors demonstrates performance suitable for low-noise quantum readout. Dark current measurements performed at a reverse bias of $-3$~V show consistently sub-nanoampere dark currents across all photodiode geometries, with no statistically meaningful dependence on detector design, indicating that the multi-bounce absorption strategy does not introduce excess current leakage. Photocurrent measurements were carried out using calibrated on-chip reference photodiodes to enable relative comparison across device geometries, followed by absolute responsivity calibration. We measure a maximum quantum efficiency of $99^{+1}_{-12}$\% (bounded) for the circular multi-bounce photodiode design. The uncertainty is dominated by the relative standard error in the lensed fiber-to-waveguide coupling rate ($11$\%) and the responsivity of the power meter used to calibrate the injected power ($5$\%) (see supplementary information).

An integrated balanced homodyne detector was constructed with a differential balanced pair of circular multi-bounce photodiodes and a custom low-noise transimpedance amplifier (TIA) made in a 90~nm SiGe process, as shown in Fig. \ref{Fig4}b. The TIA uses a shunt feedback topology with 5.9~k$\Omega$ of transimpedance gain, including an integrated diplexer to separate the buffered squeezed light detection signal and the MZI angle control signal. Placing the diplexer after the transimpedance stage prevents amplification of its added noise, which improves the signal-to-noise ratio of both the low and high frequency outputs.

The 3-dB bandwidth of the co-packaged photoreceiver was measured to be 510~MHz, with a minimum equivalent input-referred current noise of 1.9~pA/$\sqrt{\mathrm{Hz}}$ within this bandwidth as shown in Figure \ref{Fig4}c. Shot noise clearance measurements were taken up to 7~mA of photocurrent where \textgreater30 dB of shot noise clearance is seen from 10~MHz to \textgreater1 GHz. This photoreceiver exhibits a maximum measured CMRR of 50.4 dB at 10 MHz without the use of active balancing techniques (see supplementary information). These results demonstrate an extremely low-noise photoreceiver with the potential to detect squeezing ratios up to 30~dB over a wide range of frequencies. Together, these results establish a heterogeneous InP-on-SiN photodetector platform that combines near-unity on-chip quantum efficiency with low dark current, providing a key enabling component for high-efficiency, low-noise quantum photonic readout.

\section*{Discussion and Further Integration}

Achieving full heterogeneous integration of all essential photonic and electronic components on a single chip is a central goal for compact quantum photonic systems, e.g., on-chip squeezed light generation and detection. Figure 5a illustrates this vision, where integrated III-V lasers provide low-noise pump and local oscillator fields that are routed through ultra-low-loss SiN circuits to drive InGaP-based SHG and OPO stages for on-chip squeezing. Tunable interferometers and switches enable reconfigurable routing and input/output, while integrated InP photodetectors and balanced homodyne receivers provide high-efficiency, low-noise measurement. Co-integration with proximal electronics further reduces parasitics and interface losses, unifying these functionalities into a compact, phase-stable, and manufacturable quantum photonic transceiver architecture.

The component-level measurements performed in this work allow us to define bounds on measurable squeezing of the fully-integrated system. For the detection path alone, we consider four dominant degradation effects: (i)~Quantum efficiency of the integrated photodiodes, (ii)~dark noise of the integrated balanced photoreceiver, (iii)~relative-intensity noise (RIN) of the local oscillator, and (iv)~laser frequency noise. We consider a $1$~mW local oscillator and normalize all noise spectra to this relative shot noise level (see supplementary information). Figure~5b shows the maximum measurable squeezing versus proximity to threshold for an assumed local oscillator RIN of $-150$~dB/Hz, where negligible degradation occurs for the high CMRR of the integrated MZI. We then fix the proximity to threshold and observe the impact of higher RIN, where significant degradation occurs only for RIN levels greater than $-120$~dB/Hz ($30$~dB worse RIN than that of the lasers measured in this work). Phase noise degrades the measured squeezing level by partially projecting the anti-squeezed quadrature onto the squeezed quadrature \cite{Aoki2006}. Laser frequency noise can couple as phase noise due to optical path-length differences. For the linewidths achievable with the presented tunable semiconductor lasers combined with the ability to accurately match the optical path lengths in the integrated circuit, the corresponding rms phase noise contribution can be estimated in the $\upmu$rad range or below, yielding a negligible squeezing degradation. We conclude that the limiting degradation mechanism of the detection system will likely be the quantum efficiency of the photodiodes. 

Looking forward, the platform demonstrated here establishes a clear roadmap toward fully integrated, large-scale quantum photonic systems in which sources, nonlinear processors, routing networks, and detectors are seamlessly combined with control electronics on a common wafer. Continued advances in interlayer coupling, loss reduction, and co-design of photonic-electronic subsystems will enable higher levels of circuit complexity, improved noise performance, and increased functionality, including multiplexed quantum state generation, adaptive measurement, and real-time feedback. Beyond nonlinear generation and squeezing, this architecture can be extended to support frequency conversion across disparate quantum systems, hybrid integration with emerging materials, and deployment in fieldable quantum sensing and communication systems. By bridging the gap between high-performance discrete components and scalable manufacturing, heterogeneous III-V-on-SiN photonics provides a compelling foundation for realizing practical, low-noise, and fully integrated quantum photonic technologies.

\vspace{10pt}
\section*{Funding}
\noindent This work was supported by the Defense Advanced Research Projects Agency (Award No. D24AC00166-00) and the NSF (Quantum Foundry Grant No. DMR-1906325, NSF CAREER Program Grant No. 2045246, and NRT Training Program Grant No. 2152201). L.T. and L.W. acknowledge support from the NSF Graduate Research Fellowship Program. M.A.G. was supported by an appointment to the Intelligence Community Postdoctoral Research Fellowship Program at the Massachusetts Institute of Technology, administered by Oak Ridge Institute for Science and Education (ORISE) through an interagency agreement between the U.S. Department of Energy and the Office of the Director of National Intelligence (ODNI)
\section*{Acknowledgment}
\noindent A portion of this work was performed in the UCSB Nanofabrication Facility, an open access laboratory. The authors gratefully acknowledge Daniil Lukin (Brightlight Photonics) for MZI and waveguide facet polishing. The authors gratefully acknowledge the broader Nexus Photonics team, including members who are not listed as co-authors, for their contributions to the development, fabrication, testing, and continuous improvement of the heterogeneous photonic integration platform that supported this work.
\section*{Disclosures}
\noindent The authors declare no conflicts of interest.
\section*{Data Availability}
\noindent The data that support the figures in this paper and other findings of this study are available from the corresponding author on reasonable request.


\thispagestyle{plain}




\thispagestyle{plain}

\def\bibsection{\section*{References}}
\bibliography{references.bib}

\thispagestyle{plain}

\end{document}